\documentclass[runningheads]{llncs}
\usepackage[T1]{fontenc}
\usepackage{graphicx,verbatim}
\usepackage{url}
\usepackage{graphicx}
\usepackage{booktabs}
\usepackage{multirow}
\usepackage{colortbl}
\usepackage{xcolor}
\usepackage{amsmath,amssymb,graphicx}
\usepackage{bm}
\usepackage{bbm}
\usepackage{graphicx, wrapfig}
\definecolor{customblue}{RGB}{75,75,225}
\usepackage{makecell}
\usepackage{booktabs}
\usepackage{multirow}
\usepackage{amssymb}
\usepackage{subcaption}
\usepackage{kotex}

\newcommand{\ourmodel}{TISC}

\begin{document}

\def\thefootnote{$^{\dagger}$}\footnotetext{Correponding author}

\title{Anatomically Consistent TMJ Disc Segmentation via Semantic Anchoring and Clinical Priors}

\titlerunning{Anatomically Consistent TMJ Disc Segmentation}

\author{Dayun Ju\inst{1}\quad  
Chanyoung Kim\inst{1,2}\quad  
Sunyoung Jung\inst{1}\quad  
Hyo-Jung Jung\inst{3}\\  
Chena Lee\inst{3}\quad  
Younjung Park\inst{3}\quad  
Seong Jae Hwang$^{\dagger}$\inst{1}}  

\authorrunning{D. Ju et al.}

\institute{Department of Artificial Intelligence, Yonsei University, Seoul, Republic of Korea \and
Department of Computer Science, Emory University, Atlanta, Georgia, USA \and
College of Dentistry, Yonsei University, Seoul, Republic of Korea\\
\email{\tt\small \{juda0707, chanyoung, sunyoungj, seongjae\}@yonsei.ac.kr,\\ 
chanyoung.kim@emory.edu,\\ 
\tt\small \{hjjung, chenalee, darkstar\}@yuhs.ac}
}

\maketitle
\begin{abstract} 

Segmenting the temporomandibular joint (TMJ) disc from MRI is essential for accurate diagnosis of internal derangement, yet it remains unreliable in practice due to its small size, low contrast, and morphological variability. Existing methods, primarily adapted from general segmentation architectures, often produce fragmented or anatomically inconsistent masks, leading to unstable measurements of disc position and shape for downstream diagnosis.
To address these challenges, we propose \ourmodel, a TMJ disc segmentation framework that integrates semantic anchoring with clinical metadata-guided boundary refinement. 
The framework first establishes robust disc localization in the foundation model feature space via a Prototypical Semantic Anchoring (PSA) module that aggregates adjacent-slice MedDINOv3 features and derives a prototype-driven similarity map. 
It then performs targeted boundary refinement through a Clinical-Metadata Point Refinement (C-MPR) module, with point-wise predictions modulated by Mouth Open Limitation (MOL), a clinical indicator associated with disc displacement without reduction.
On a large-scale cohort of 2,488 PD MRI volumes from 1,300 patients, our method achieves up to a 4.96 Dice improvement over strong baselines across diverse architectures, delivering more anatomically coherent and clinically reliable TMJ disc segmentation.


\end{abstract}
\section{Introduction}

Temporomandibular joint (TMJ) disc is a vital fibrocartilaginous structure that ensures smooth mandibular movement and optimal load distribution~\cite{okeson2019management}. Accurate interpretation of the disc is essential for managing internal derangement, which affects up to 12\% of the population and is a leading cause of chronic orofacial pain~\cite{mol_3}. 
Although Magnetic Resonance Imaging (MRI) is the gold standard for diagnosis, reliable assessment of disc displacement and structural abnormalities remains challenging, often requiring highly specialized radiological expertise to resolve ambiguities between the disc and surrounding soft tissues~\cite{bag2014imaging,larheim2015temporomandibular}.
If automated segmentation were available, it could facilitate objective and reproducible quantification of disc displacement and morphology, enabling more standardized treatment planning.
However, this task is particularly difficult, as the disc typically occupies only about 0.2\% of the total pixel area on an MRI slice and exhibits low contrast relative to adjacent ligaments and effusions~\cite{nozawa2022automatic}.

In the TMJ domain, various data-driven approaches have been explored for diagnostic modeling and clinical assessment~\cite{tmj_seg,jung2026multimodal,tmj_seg_2}. Among these efforts, TMJ disc segmentation has primarily been addressed using general segmentation architectures, such as UNet-based models~\cite{swinunetr,unetr,nnunet,unet}. More recently, medical foundation models, including MedSAM~\cite{medsam} have also been applied to medical image segmentation.
Despite their broad utility, general-purpose segmentation models struggle in TMJ disc segmentation for two primary reasons. First, reliable localization is inherently difficult because the disc is extremely small and exhibits low contrast, providing limited semantic cues~\cite{nozawa2022automatic}. This problem becomes more pronounced in pathological conditions such as disc displacement without reduction, where the disc deviates from its typical anatomical position. 
Consequently, spatial uncertainty increases, weakening positional cues and making models prone to confusion with adjacent anatomy or imaging artifacts, leading to false positives. 
Second, even when roughly localized, precise boundary delineation remains challenging. The disc lacks a consistent shape and may appear folded, compressed, or morphologically distorted across disease states, while subtle intensity transitions with neighboring tissues further obscure its margins. These factors often result in ambiguous boundaries and over-segmentation into surrounding structures~\cite{nozawa2022automatic}.

Motivated by these challenges, we propose \ourmodel, a TMJ disc segmentation framework that integrates semantic anchoring with metadata-guided boundary refinement. 
The framework first establishes disc localization in the foundation model feature space. To this end, the Prototypical Semantic Anchoring (PSA) module aggregates adjacent-slice MedDINOv3~\cite{meddinov3} features and employs a cross-attention query extractor to generate a similarity-based localization map at the bottleneck, stabilizing predictions and reducing false positives under pathological displacement. 
Building upon this disc-focused representation, the Clinical-Metadata Point Refinement (C-MPR) module adaptively adjusts boundary predictions at the point level~\cite{pointrend} conditioned on clinical metadata, enabling correction of ambiguous regions. 
We further incorporate Mouth Open Limitation (MOL), a binary indicator of limited mouth opening (<40\,mm)~\cite{mol_3}, as auxiliary metadata due to its strong association with disc displacement without reduction (DDwoR), a condition characterized by abnormal disc position and deformation~\cite{mol_place_1,mol_shape_2,mol_place_2,mol_1_shape}. Treated as an indicator of displacement severity, MOL guides spatial attention in PSA and modulates refinement features via FiLM layers, promoting adaptive localization and boundary correction across pathological states.

The proposed framework is evaluated on a large-scale in-house cohort comprising 2,488 PD MRI volumes and 4,601 manually annotated slices from 1,300 patients.
Through extensive patient-level cross-validation, our model consistently outperforms state-of-the-art baselines across diverse architectures. It achieves notable improvements in both region-level and boundary metrics, producing sharper and anatomically coherent segmentation.

\section{Method}

\begin{figure}[t]
\centering
\includegraphics[width=\linewidth]{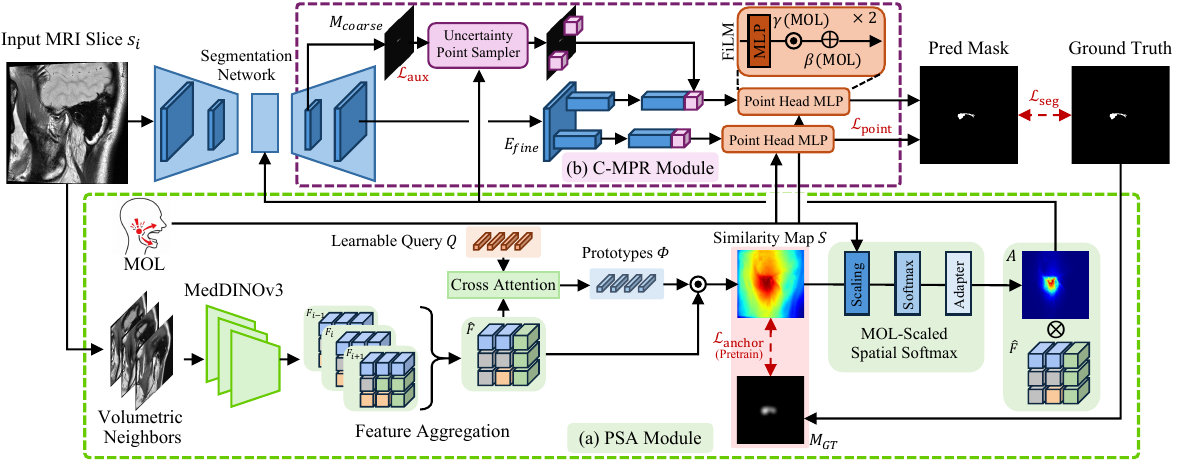}  
\caption{
Overview of \ourmodel.
(a) The PSA module establishes disc localization at the bottleneck, followed by (b) the C-MPR module for boundary refinement during upscaling, enabling anatomically consistent and precise segmentation. 
}
\label{fig:overview}
\end{figure}

As illustrated in Fig.~\ref{fig:overview}, we propose \ourmodel~ for TMJ disc segmentation via semantic anchoring and clinical metadata-guided mask refinement. The framework first employs a semantic anchoring module (PSA, Sec. 2.1) that distills foundation model features to establish anatomical anchors for localization. It then incorporates a clinical metadata-guided point refinement module (C-MPR, Sec. 2.2) for fine-grained boundary refinement to ensure anatomical fidelity of the mask.

Notably, we further augment this framework by integrating Mouth Open Limitation (MOL), defined as mouth opening less than 40 mm, metadata as a clinical surrogate for disc displacement without reduction (DDwoR). Clinically, a positive MOL status (MOL=1) is strongly associated with DDwoR~\cite{mol_1_shape,mol_3}, a condition characterized by abnormal disc position~\cite{mol_place_1,mol_place_2} and exhibiting deformed morphologies such as folded or globular shapes~\cite{mol_shape_2}. We leverage these insights by using MOL to adaptively scale attention temperature in the PSA module for wider spatial anchoring and to guide feature modulation via FiLM layers in the C-MPR module. This design enables the model to capture structural details and maintain anatomical consistency across diverse disease states.

\subsection{Disc-localized Multi-context Encoding}
To exploit high-fidelity foundation model features for anatomical structures, we propose the Prototypical Semantic Anchoring (PSA) module. PSA reinforces disc signals via (I) volumetric context and (II) distills anatomical identity through a learnable query in the MedDINOv3~\cite{meddinov3} space. 
(III) Projecting this identity back to the spatial domain establishes a precise anchor to guide segmentation.

\begin{figure}[t]
  \centering
  \includegraphics[width=\linewidth]{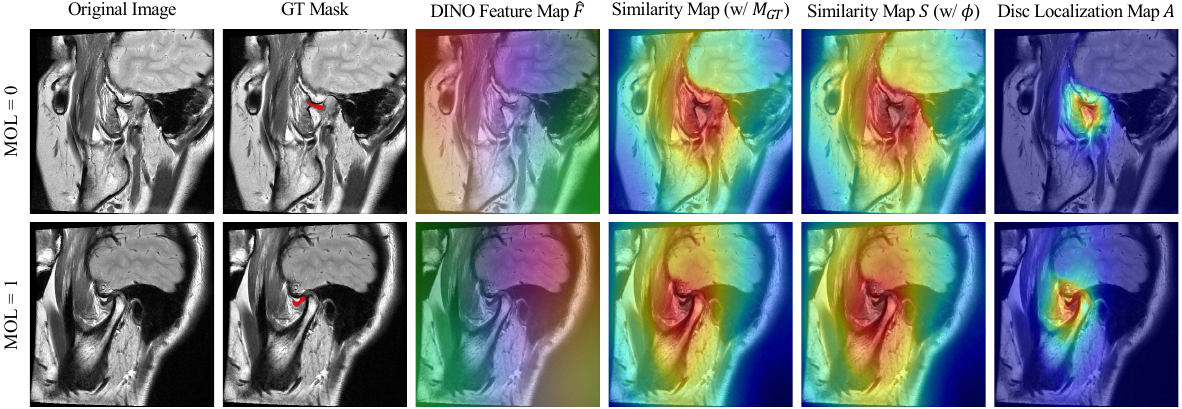}  
  \caption{
  Visualization of similarity-based disc localization maps. From left to right, the figure shows the original MRI slice, the ground truth disc mask, the MedDINOv3 feature map, the similarity map using $M_{GT}$, the similarity map from the learned prototype, and the final localization map after temperature scaling.
  }
  \label{fig:feature_map}
\end{figure}

\subsubsection{(I) Multi-context Feature Aggregation.}
To incorporate volumetric context, we aggregate features from a target slice $s_i$ and its adjacent slices $s_{i\pm1}$. Each slice $s_j$ ($j \in \{i-1, i, i+1\}$) is encoded by a pre-trained MedDINOv3 encoder to obtain feature maps $F_j \in \mathbb{R}^{L \times C}$, where $L$ denotes the number of spatial tokens and $C$ the embedding dimension. We then compute an aggregated representation as $\hat{F} = (F_{i-1} + 2F_i + F_{i+1})/4$.
This aggregation enforces inter-slice consistency and stabilizes disc representations against slice-level ambiguity and noise.

\subsubsection{(II) Semantic Prototype Extraction.}
Using the aggregated feature map $\hat{F}$, the PSA module distills the disc-representative semantic features into a latent prototype $\Phi$
to determine \textit{what} constitutes the target, providing a semantic foundation for modulating the UNet bottleneck.
To achieve this, we define a single learnable query $Q \in \mathbb{R}^{C}$ that selectively aggregates spatial information via a cross-attention mechanism:
$\Phi = \text{Softmax}\left(\frac{Q \hat{F}^\top}{\sqrt{C}}\right) \hat{F}$,  where $\hat{F}$ acts as a key and value.
This operation selectively pools features across the spatial dimension $L$ to form a latent prototype that contains the disc’s semantic properties.
To spatially align the prototype $\Phi$, we optimize PSA with a spatial alignment loss
$\mathcal{L}_{anchor} = \text{BCE}(\text{Norm}(S), M_{GT}),$
where $\text{BCE}$ denotes the Binary Cross Entropy loss. Here, $S$ is the cosine similarity map between $\Phi$ and the aggregated feature map $\hat{F}$, and $\text{Norm}(\cdot)$ normalizes $S$ to $[0,1]$. The ground truth mask $M_{GT}$ is Gaussian-smoothed to prevent overly sharp supervision and encourage spatial tolerance.
This ensures the query-based filter remains focused on the disc’s identity even under complex imaging artifacts.

\subsubsection{(III) Disc Localization and Bottleneck Modulation.} 

Given the similarity map $S$, we project the distilled semantic identity onto the spatial feature grid to determine where feature modulation should occur. 
We obtain a Disc Localization Map $A$ by applying a temperature-scaled spatial softmax to $S$ as 
\begin{equation}
A(y,x) =
\frac{\exp\left(S(y,x)/\tau\right)}
{\sum_{y',x'} \exp\left(S(y',x')/\tau\right)},
\end{equation}
where $(y,x)$ are spatial indices.
The temperature $\tau$ is modulated by MOL, as illustrated in Fig.~\ref{fig:feature_map}. 
Specifically, $\tau=0.05$ for MOL=0 yields a sharper attention map over stable anatomical regions, whereas $\tau=0.1$ produces a broader distribution to accommodate pathological displacement, reflecting the higher likelihood of disc displacement when MOL=1.
The resulting map $A$ acts as a semantic spatial mask that modulates $\hat{F}$ via element-wise multiplication, suppressing non-disc responses. The localized features are then injected into the UNet bottleneck through an additive skip connection to guide decoding toward disc-relevant regions. 

\subsection{Metadata-Augmented Point Refinement}
To enable fine-grained adjustments at the point level~\cite{pointrend}, we introduce a Clinical-Metadata-Guided Point Refinement (C-MPR) module. By leveraging MOL as a clinical prior, C-MPR reduces morphological ambiguities and enables high-fidelity boundary reconstruction, particularly in cases of severe pathological displacement. This module operates through two main stages: (I) Progressive Point-based Upsampling and (II) Metadata-Driven Feature Modulation.

\subsubsection{(I) Progressive Point-based Upsampling.} 

The C-MPR module performs boundary refinement starting from a coarse logit map 
$M_{coarse} \in \mathbb{R}^{1 \times H/2 \times W/2}$ predicted by the decoder. 
To preserve global structural coherence, $M_{coarse}$ is supervised with an auxiliary loss 
$\mathcal{L}_{aux}$ combining BCE and Soft Dice.
Unlike existing uncertainty-driven refinement~\cite{pointrend}, C-MPR adopts a semantic-guided point sampling strategy. 
We sample $P$ candidate points by combining (i) uncertainty-based selection near the decision boundary, where coarse probabilities approach 0.5, and (ii) prototype-aware sampling from regions with high similarity responses in $S$. 
Approximately 25\% of points are drawn from semantically salient regions, promoting refinement around disc-relevant structures.
For each selected point, a feature vector $h$ is constructed by concatenating coarse logits with high-resolution encoder features $E_{fine} \in \mathbb{R}^{C_e \times H \times W}$. 
This design enables localized boundary correction guided by prediction uncertainty and semantic identity.

\subsubsection{(II) Metadata-Driven Feature Modulation.}

To handle morphology changes under pathological conditions, we process point-wise features $h$ using a FiLM-based MLP~\cite{film} conditioned on MOL.
While PSA provides a global anatomical anchor at the bottleneck, $h$ captures local multi-scale context, enabling C-MPR to adapt boundary refinement based on clinical metadata.
Each FiLM layer applies a channel-wise affine transformation
\begin{equation}
h' = \gamma(\text{MOL}) \odot h + \beta(\text{MOL}),
\end{equation}
where $\gamma(\text{MOL})$ and $\beta(\text{MOL})$ denote scaling and shifting vectors generated from a learnable embedding of MOL.
This conditioning adaptively reweights feature responses, increasing sensitivity to structural deformations in MOL=1 cases while maintaining stable boundary delineation in MOL=0 cases.
The refined predictions are supervised with a point-wise BCE loss $\mathcal{L}_{point}$.

\subsubsection{Total Loss.}
The framework is optimized via a multi-task objective function:
\begin{equation}
    \mathcal{L}_{total} = \mathcal{L}_{seg} + {\lambda}_{point}\mathcal{L}_{point} + {\lambda}_{aux}\mathcal{L}_{aux}.
\end{equation}
where $\mathcal{L}_{seg}$ combines BCE and soft dice loss. The weighting factors ${\lambda}_{point}$ and ${\lambda}_{aux}$ regulate the contributions of $\mathcal{L}_{point}$ and $\mathcal{L}_{aux}$, respectively.

\section{Experiments}

\begin{table}[t]
\centering
\caption{ 
Quantitative Analysis.
Results are reported as mean(standard deviation) over 5-fold cross-validation.
Performance is evaluated using Dice, clDice, Detection rate, HD95, and ASSD.
$^\dagger$ indicates $p < 0.05$ compared to baselines.
}
\label{table:tab_1_main}
\footnotesize
\setlength{\tabcolsep}{4.1pt}
\begin{tabular}{lc ccccc}
\toprule[1.2pt]
\multirow{2.4}{*}{\textbf{Method}} & \multicolumn{3}{c}{Region-level (\%) $\uparrow$} & & \multicolumn{2}{c}{Boundary (mm) $\downarrow$} \\
\cmidrule(lr){2-4} \cmidrule(lr){6-7}
& Dice & clDice & Detection & & HD95 & ASSD \\
\midrule
UNETR~\cite{unetr} & 58.96(0.38) & 63.58(0.71) & 81.66(1.59) & & 13.51(0.40) & 2.96(0.07) \\
MedSAM~\cite{medsam} & 53.30(0.51) & 63.11(0.94) & 69.73(2.01) & & 10.33(0.07) & 2.68(0.03) \\
\midrule
UNet$^\dagger$~\cite{unet} & 76.58(0.33) & 86.23(0.70) & 97.41(1.22) & & 7.19(0.38) & 1.37(0.07) \\
\rowcolor{gray!15}
$+$ Ours & 80.46(0.36) & 89.94(0.58) & 98.85(0.38) & & 5.23(0.16) & 1.02(0.04) \\
\midrule
nnUNet$^\dagger$~\cite{nnunet} & 78.02(0.22) & 87.54(0.33) & 98.41(0.53) & & 6.33(0.34) & 1.23(0.03) \\
\rowcolor{gray!15}
$+$ Ours & 80.12(0.96) & 89.77(1.14) & 98.62(0.45) & & 5.17(0.35) & 1.27(0.28) \\
\midrule
SwinUNETR$^\dagger$~\cite{swinunetr} & 75.34(0.41) & 84.21(0.62) & 96.52(0.82) & & 7.69(0.25) & 1.57(0.05) \\
\rowcolor{gray!15}
$+$ Ours & 80.30(0.35) & 89.67(0.54) & 99.00(0.27) & & 5.16(0.24) & 1.03(0.07) \\
\bottomrule[1.2pt]
\end{tabular}
\end{table}

\subsection{Experimental Settings}

\noindent \textbf{Dataset and Evaluation Metrics.}
The proposed framework is evaluated on a private dataset of 2,488 PD-weighted MRI volumes from 1,300 patients, following Institutional Review Board approval and the Declaration of Helsinki~\cite{helsinki}. Images are acquired with a 3.0 T GE Pioneer scanner using specific parameters: TE/TR of 50/2084 ms, a 130$\times$130 mm FOV, and a 2.5 mm slice thickness. 
For ground truth segmentation masks, 4,601 sagittal slices are annotated and verified through specialist consensus.
The dataset also includes patient-level MOL and expert-annotated disc displacement labels.
We conduct patient-level 5-fold cross-validation to ensure robustness and prevent data leakage. Performance was evaluated using Dice~\cite{dice}, clDice~\cite{cldice}, and Detection (IoU > 0.3) for region-level performance, along with 95th percentile Hausdorff Distance (HD95) and Average Symmetric Surface Distance (ASSD) to measure boundary accuracy.

\noindent \textbf{Implementation Details.}
To ensure reproducibility and consistency, a uniform hyperparameter configuration was maintained across all cross-validation folds. 
The weighting coefficients ${\lambda}_{point}$ and ${\lambda}_{aux}$ are set to 20 and 0.8, respectively.
The PSA module was pre-trained for 20 epochs and subsequently frozen during segmentation training. 
For the full pipeline, a warm-up was performed for 10 epochs to stabilize the encoder before activating the C-MPR module. All images and corresponding labels are used at their original resolution of $512 \times 512$.

\begin{figure}[t]
  \centering
  \includegraphics[width=\linewidth]{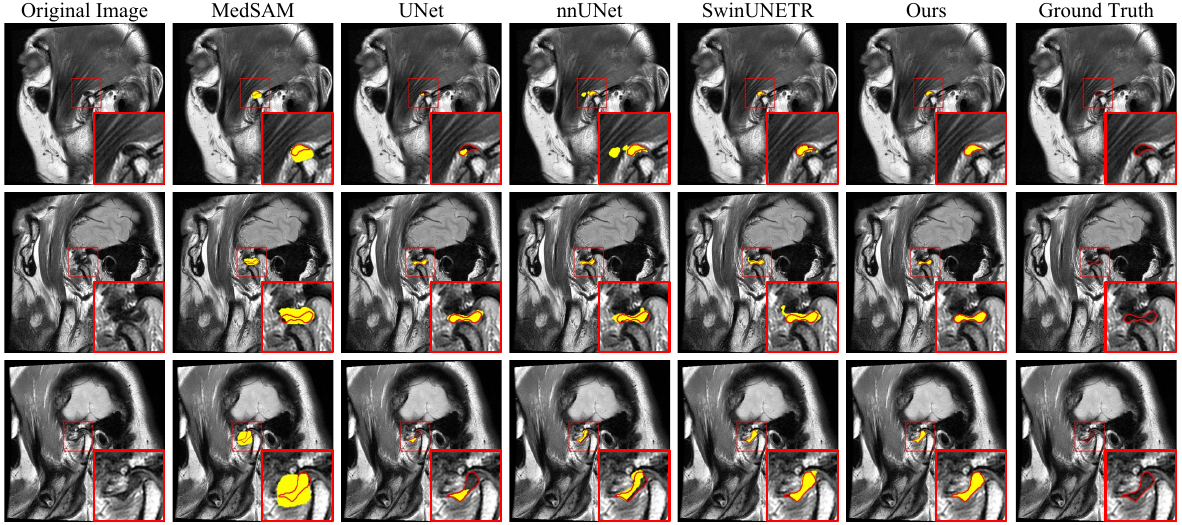}  
  \caption{
  Qualitative Results. Segmentation results of our model are compared against MedSAM~\cite{medsam}, UNet~\cite{unet}, nnUNet~\cite{nnunet}, and SwinUNETR~\cite{swinunetr}. 
  Yellow regions represent predicted masks, and the red boundary denotes the ground truth.
  }
  \label{fig:qualitative}
\end{figure}

\subsection{Evaluation Results}

\subsubsection{Quantitative Evaluation.}
Table \ref{table:tab_1_main} presents the average 5-fold cross-validation results. 
Our framework consistently improves both region-level and boundary metrics over state-of-the-art baselines. When integrated with UNet~\cite{unet}, the proposed method achieves the highest Dice score of 80.46\%, outperforming the original UNet (76.58\%). Substantial improvements are also observed in boundary metrics, with HD95 reduced by 32.9\% on the SwinUNETR baseline~\cite{swinunetr} (7.69 to 5.16). 
These results highlight its robustness across diverse segmentation architectures, consistently improving both localization and boundary precision.

\noindent \textbf{Qualitative Analysis.}
Fig.~\ref{fig:qualitative} presents representative qualitative comparisons between our framework (UNet-based) and baseline models. The results show that baseline methods frequently produce partial segmentations that capture only the anterior portion of the disc or exhibit oversegmentation beyond the ground truth boundary (red contour). In contrast, our framework more accurately delineates the complete disc morphology, closely adhering to the annotated boundary and preserving structural continuity across the thin intermediate zone.


\begin{table*}[t]
\centering
\caption{Ablation study on PSA, C-MPR components and MOL metadata.}
\label{tab:ablation}

\centering
\vspace{0pt}
{\small
\setlength{\tabcolsep}{8.1pt}
\begin{tabular}{ccccccc}
\toprule
\multirow{2}{*}{Exp.\#} 
& \multicolumn{2}{c}{PSA} 
& \multirow{2}{*}{C-MPR} 
& \multirow{2}{*}{MOL} 
& \multirow{2}{*}{Dice $\uparrow$} 
& \multirow{2}{*}{HD95 $\downarrow$} \\
\cmidrule(lr){2-3}
& Prototype & Vol. Neighbor & & & & \\
\midrule
1 &  &  &  &  & 76.54 & 7.49 \\
2 & \checkmark &  &  & \checkmark & 78.99 & 6.22 \\
3 & \checkmark & \checkmark &  & \checkmark & 79.55 & 6.01 \\
4 &  &  & \checkmark & \checkmark & 78.75 & 5.30 \\
5 & \checkmark & \checkmark & \checkmark &  & 78.65 & 6.35 \\
\rowcolor{gray!15}
6 & \checkmark & \checkmark & \checkmark & \checkmark & 80.10 & 5.27 \\
\bottomrule
\end{tabular}
}
\hspace{0.03\textwidth}

\end{table*}

\subsection{Ablation Study}
Table~\ref{tab:ablation}(a) shows ablation study results.
Given the consistent trends across all validation folds (Sec. 3.2), 
we conduct ablation experiments on Fold 0 using a standard UNet~\cite{unet} backbone to isolate the contribution of each module.

\noindent \textbf{Effect of PSA.}
Exp. 2 shows that incorporating PSA with single-slice features ($F$) improves Dice from 76.54 to 78.99. Extending PSA to multi-slice contextual features ($\hat{F}$) in Exp. 3 further increases Dice to 79.55. These results indicate that multi-context PSA enhances disc localization.

\noindent \textbf{Effect of C-MPR.}
Comparing Exp. 1 and Exp. 4 highlights the contribution of the C-MPR module. Incorporating C-MPR reduces HD95 from 7.49 to 5.30 (29.2\% reduction) while improving Dice from 76.54 to 78.75, demonstrating enhanced boundary accuracy. These results indicate that metadata-guided point refinement improves structural continuity.

\noindent \textbf{Effect of Clinical Metadata Conditioning.}
Removing MOL-guided conditioning from PSA and C-MPR (Exp. 5) results in a performance drop compared to the full model (Exp. 6). This indicates that incorporating MOL provides meaningful clinical cues that enhance robustness under pathological variability.

\subsection{Application}
\noindent \textbf{Segmentation-Assisted Classification.}
To assess whether segmentation masks improve downstream diagnosis, we performed binary classification of normal versus disc displacement.
Close–open MRI slice pairs from the same joint side are used as a 2-channel input to a ResNet18~\cite{resnet} baseline. 
Adding predicted masks expanded it to a 4-channel input. The vanilla model achieved an AUROC of 0.61, which increased to 0.73 with our masks, approaching the performance obtained using ground truth masks (0.74). These findings indicate that accurate disc segmentation provides meaningful structural cues for downstream diagnosis.

\section{Conclusion}

We present \ourmodel, a framework for anatomically consistent TMJ disc segmentation in PD-weighted MRI that unifies semantic anchoring with clinically guided refinement. The PSA module establishes a stable localization prior in the foundation model feature space via multi-slice aggregation and prototype-based modulation, mitigating mislocalization under pathological displacement, while the C-MPR module leverages MOL metadata to preserve structural continuity in morphologically distorted cases. Extensive patient-level validation confirms consistent improvements over strong baselines. By tightly coupling foundation model semantics with clinically meaningful priors, the framework enables reliable segmentation of small, low-contrast, pathology-sensitive structures.
\begin{credits}

\subsubsection{\ackname} This work was supported in part by the IITP RS-2024-00457882 (AI Research Hub Project), IITP 2020-II201361, NRF RS-2024-00345806, NRF RS-2023-002620, and RQT-25-120390.


\end{credits}

\bibliographystyle{splncs04}
\bibliography{main}

@string{cvpr      = {Proceedings of the IEEE/CVF Conference on Computer Vision and Pattern Recognition}}

@string{aaai      = {Proceedings of the Conference on Artificial Intelligence}}

@string{wacv      = {Proceedings of the IEEE Winter Conference on Applications of Computer Vision}}

@string{miccai    = {Proceedings of the International Conference on Medical Image Computing and Computer Assisted Intervention}}

@string{cvpr     = {CVPR}}

@string{aaai     = {AAAI}}

@string{wacv     = {WACV}}

@string{miccai   = {MICCAI}}

@inproceedings{unet,
  title={U-net: Convolutional networks for biomedical image segmentation},
  author={Ronneberger, Olaf and Fischer, Philipp and Brox, Thomas},
  booktitle=miccai,
  pages={234--241},
  year={2015},
  organization={Springer}
}

@article{nnunet,
  title={nnu-net: Self-adapting framework for u-net-based medical image segmentation},
  author={Isensee, Fabian and Petersen, Jens and Klein, Andre and Zimmerer, David and Jaeger, Paul F and Kohl, Simon and Wasserthal, Jakob and Koehler, Gregor and Norajitra, Tobias and Wirkert, Sebastian and others},
  journal={arXiv preprint arXiv:1809.10486},
  year={2018}
}

@inproceedings{swinunetr,
  title={Swin unetr: Swin transformers for semantic segmentation of brain tumors in mri images},
  author={Hatamizadeh, Ali and Nath, Vishwesh and Tang, Yucheng and Yang, Dong and Roth, Holger R and Xu, Daguang},
  booktitle={International MICCAI brainlesion workshop},
  pages={272--284},
  year={2021},
  organization={Springer}
}

@article{meddinov3,
  title={MedDINOv3: How to adapt vision foundation models for medical image segmentation?},
  author={Li, Yuheng and Wu, Yizhou and Lai, Yuxiang and Hu, Mingzhe and Yang, Xiaofeng},
  journal={arXiv preprint arXiv:2509.02379},
  year={2025}
}

@article{medsam,
  title={Segment Anything in Medical Images},
  author={Ma, Jun and He, Yuting and Li, Feifei and Han, Lin and You, Chenyu and Wang, Bo},
  journal={Nature Communications},
  volume={15},
  pages={654},
  year={2024}
}

@inproceedings{pointrend,
  title={Pointrend: Image segmentation as rendering},
  author={Kirillov, Alexander and Wu, Yuxin and He, Kaiming and Girshick, Ross},
  booktitle=cvpr,
  pages={9799--9808},
  year={2020}
}

@inproceedings{cldice,
  title={clDice-a novel topology-preserving loss function for tubular structure segmentation},
  author={Shit, Suprosanna and Paetzold, Johannes C and Sekuboyina, Anjany and Ezhov, Ivan and Unger, Alexander and Zhylka, Andrey and Pluim, Josien PW and Bauer, Ulrich and Menze, Bjoern H},
  booktitle=cvpr,
  pages={16560--16569},
  year={2021}
}

@article{dice,
  title={Measures of the amount of ecologic association between species},
  author={Dice, Lee R},
  journal={Ecology},
  volume={26},
  number={3},
  pages={297--302},
  year={1945},
  publisher={JSTOR}
}

@article{helsinki,
  title={World Medical Association Declaration of Helsinki: ethical principles for medical research involving human subjects},
  author={World Medical Association and others},
  journal={Jama},
  volume={310},
  number={20},
  pages={2191--2194},
  year={2013},
  publisher={American Medical Association}
}

@article{TMJ_seg,
  title={Automated segmentation of articular disc of the temporomandibular joint on magnetic resonance images using deep learning},
  author={Ito, Shota and Mine, Yuichi and Yoshimi, Yuki and Takeda, Saori and Tanaka, Akari and Onishi, Azusa and Peng, Tzu-Yu and Nakamoto, Takashi and Nagasaki, Toshikazu and Kakimoto, Naoya and others},
  journal={Scientific Reports},
  volume={12},
  number={1},
  pages={221},
  year={2022},
  publisher={Nature Publishing Group UK London}
}

@article{TMJ_seg_2,
  title={Temporomandibular joint segmentation in MRI images using deep learning},
  author={Li, Mengxun and Punithakumar, Kumaradevan and Major, Paul W and Le, Lawrence H and Nguyen, Kim-Cuong T and Pacheco-Pereira, Camila and Kaipatur, Neelambar R and Nebbe, Brian and Jaremko, Jacob L and Almeida, Fabiana T},
  journal={Journal of Dentistry},
  volume={127},
  pages={104345},
  year={2022},
  publisher={Elsevier}
}

@inproceedings{unetr,
  title={Unetr: Transformers for 3d medical image segmentation},
  author={Hatamizadeh, Ali and Tang, Yucheng and Nath, Vishwesh and Yang, Dong and Myronenko, Andriy and Landman, Bennett and Roth, Holger R and Xu, Daguang},
  booktitle=wacv,
  pages={574--584},
  year={2022}
}

@article{mol_1_shape,
  title={Long-term changes in clinical signs and symptoms and disc position and morphology in patients with nonreducing disc displacement in the temporomandibular joint},
  author={Sato, Shuichi and Sakamoto, Maya and Kawamura, Hiroshi and Motegi, Katsutoshi},
  journal={Journal of oral and maxillofacial surgery},
  volume={57},
  number={1},
  pages={23--29},
  year={1999},
  publisher={Elsevier}
}

@article{mol_3,
  title={Diagnostic criteria for temporomandibular disorders (DC/TMD) for clinical and research applications: recommendations of the International RDC/TMD Consortium Network and Orofacial Pain Special Interest Group},
  author={Schiffman, Eric and Ohrbach, Richard and Truelove, Edmond and Look, John and Anderson, Gary and Goulet, Jean-Paul and List, Thomas and Svensson, Peter and Gonzalez, Yoly and Lobbezoo, Frank and others},
  journal={Journal of oral \& facial pain and headache},
  volume={28},
  number={1},
  pages={6},
  year={2014}
}

@article{mol_place_1,
  title={Classification of temporomandibular joint internal derangement based on magnetic resonance imaging and clinical findings of 435 patients contributing to a nonsurgical treatment protocol},
  author={Hegab, Ayman F and Al Hameed, Hossam IAbd and Karam, Khaled Said},
  journal={Scientific Reports},
  volume={11},
  number={1},
  pages={20917},
  year={2021},
  publisher={Nature Publishing Group UK London}
}

@article{mol_place_2,
  title={Temporomandibular joint disc displacement with reduction: a review of mechanisms and clinical presentation},
  author={Poluha, Rodrigo Lorenzi and Canales, Giancarlo De la Torre and Costa, Yuri Martins and Grossmann, Eduardo and Bonjardim, Leonardo Rigoldi and Conti, Paulo C{\'e}sar Rodrigues},
  journal={Journal of applied oral science},
  volume={27},
  pages={e20180433},
  year={2019},
  publisher={SciELO Brasil}
}

@article{mol_shape_2,
  title={The relationship between the degree of disk displacement and ability to perform disk reduction},
  author={Kurita, Hiroshi and Ohtsuka, Akiko and Kobayashi, Hiroichi and Kurashina, Kenji},
  journal={Oral Surgery, Oral Medicine, Oral Pathology, Oral Radiology, and Endodontology},
  volume={90},
  number={1},
  pages={16--20},
  year={2000},
  publisher={Elsevier}
}

@article{jung2026multimodal,
  title={Multimodal deep learning with anatomically constrained attention for screening MRI-detectable TMJ abnormalities from panoramic images},
  author={Jung, Hyo-Jung and Ju, Dayun and Kim, Chanyoung and Hwang, Seong Jae and Lee, Chena and Park, Younjung},
  journal={npj Digital Medicine},
  year={2026},
  publisher={Nature Publishing Group UK London}
}

@inproceedings{resnet,
  title={Deep residual learning for image recognition},
  author={He, Kaiming and Zhang, Xiangyu and Ren, Shaoqing and Sun, Jian},
  booktitle={Proceedings of the IEEE conference on computer vision and pattern recognition},
  pages={770--778},
  year={2016}
}

@article{bag2014imaging,
  title={Imaging of the temporomandibular joint: An update},
  author={Bag, Asim K and Gaddikeri, Santhosh and Singhal, Aparna and Hardin, Simms and Tran, Benson D and Medina, Josue A and Cur{\'e}, Joel K},
  journal={World journal of radiology},
  volume={6},
  number={8},
  pages={567},
  year={2014}
}

@article{larheim2015temporomandibular,
  title={Temporomandibular joint diagnostics using CBCT},
  author={Larheim, TA and Abrahamsson, AK and Kristensen, MLZA and Arvidsson, LZ},
  journal={Dentomaxillofacial Radiology},
  volume={44},
  number={1},
  pages={20140235},
  year={2015},
  publisher={Oxford University Press}
}

@book{okeson2019management,
  title={Management of Temporomandibular Disorders and Occlusion-E-Book: Management of Temporomandibular Disorders and Occlusion-E-Book},
  author={Okeson, Jeffrey P},
  year={2019},
  publisher={Elsevier Health Sciences}
}

@article{nozawa2022automatic,
  title={Automatic segmentation of the temporomandibular joint disc on magnetic resonance images using a deep learning technique},
  author={Nozawa, Michihito and Ito, Hirokazu and Ariji, Yoshiko and Fukuda, Motoki and Igarashi, Chinami and Nishiyama, Masako and Ogi, Nobumi and Katsumata, Akitoshi and Kobayashi, Kaoru and Ariji, Eiichiro},
  journal={Dentomaxillofacial Radiology},
  volume={51},
  number={1},
  pages={20210185},
  year={2022},
  publisher={Oxford University Press}
}

@inproceedings{film,
  title={Film: Visual reasoning with a general conditioning layer},
  author={Perez, Ethan and Strub, Florian and De Vries, Harm and Dumoulin, Vincent and Courville, Aaron},
  booktitle=aaai,
  volume={32},
  number={1},
  year={2018}
}
\end{document}